\documentclass[a4paper,10pt]{article}
\usepackage{enumerate}
\usepackage{color}
\usepackage[utf8]{inputenc} 
\usepackage[english]{babel}
\usepackage[T1]{fontenc}
\usepackage{graphicx}
\usepackage{amsfonts,amssymb,amsmath,latexsym,amsthm}
\usepackage{textcomp}
\usepackage[pdftex]{hyperref}
\usepackage{geometry}
\DeclareMathOperator{\sech}{sech}

\title{Solitary wave collisions for Whitham-Boussinesq systems}
\author{Marcelo V. Flamarion$^{1}$ and Rosa M. Vargas-Maga{\~ n}a$^{2}$}
\date{}

\begin{document}
\maketitle
\begin{center}
{\footnotesize $^1$Unidade Acad{\^ e}mica do Cabo de Santo Agostinho, \\
UFRPE/Rural Federal University of Pernambuco, BR 101 Sul, Cabo de Santo Agostinho-PE, Brazil,  54503-900 \\
marcelo.flamarion@ufrpe.br }

\vspace{0.3cm}
{\footnotesize $^{2}$University of Edinburgh,  School of Mathematics, Edinburgh, Scotland, EH9 3FD, UK }


\end{center}


\begin{abstract} 
This work concerns  soliton-type numerical solutions for two Whitham-Boussinesq-type models. Solitary waves are computed  using an iterative Newton-type and continuation methods with high accuracy. The method allow us to compute solitary waves with large amplitude and speed close to the singular limit. These solitary waves are set as initial data  and overtaking collisions are considered for both systems. We show that both system satisfy the geometric Lax-categorization of two-soliton collision. Numerical evidences indicate that one of the systems also admits an algebraic Lax-categorization based on the ratio of the initial solitary wave amplitudes with a different range from the one predicted by Lax. However, we show that such categorization is not possible for the second system.

	\end{abstract}

\section*{Introduction}

{ Solitary waves are prototypical nonlinear structures occurring in dispersive media, which propagates through space without a change in its shape or
size at constant speed. 
The envelope of the wave has one global peak and decays far away from the peak.
  A soliton is a nonlinear solitary wave with the additional property
that the wave retains its permanent structure, even after interacting with another soliton. For example, two solitons propagating in opposite directions effectively emerge form the interaction with no change other than possible a phase shift. In other words two solitons collide elastically.

Solitary waves have garnered significant attention both for their deep, mathematical connections to completely integrable systems and their practical
utility due to their wide range of applications in fluid mechanics, nonlinear optics, geophysics, condensed
matter physics and biology see, for instance \cite{4,5,6,7}. 
 Solitary waves are not limited to exotic media and arise in many contexts, including the elevation of the surface of water and the intensity of light in optical fibers. However, the solitary waves on non-integrable systems have been less studied \cite{Dinvay2021, Linaresetal}, mainly because the analysis is much more complex and requires sophisticated and highly precise numerical tools to determine their existence, describe their shape, trace a branch of solutions, and study their propagation in different complex physical  settings.


 The manifestation of balance between dispersion and nonlinearity for the occurrence of solitary waves 
can be quite different from system to system. In particular, in the framework of fully dispersive and weakly nonlinear long water wave
models 
 such as Whitham equations, Serre equations and Whitham-Boussinesq models. Most of these models are non-integrable and defined through nonlocal operators. 
In particular,  Whitham-Boussinesq models are of considerable current interest in mathematics, physics, geophysics and coastal engineering as nonlocal extensions of well-studied dispersive shallow water wave models, such as the Korteweg-de Vries equation (KdV) and Boussinesq systems. Recent results for these models in constant depth are the existence of periodic traveling waves (Ehrnstr\"om, Kalisch, 2009) \cite{ehrnstrom2009traveling}, solitary waves (Ehrnstr\"om, Groves, Wahl\'en, 2012)   \cite{ehrnstrom2012existence}, wave breaking (Naumkin, Shishmarev, 1994) \cite{naumkin1994nonlinear}, (Constantin, Escher, 1998)  \cite{constantin1998wave} (Hur, 2015) \cite{hur2015breaking} and limiting Stokes waves (Ehrnstr\"om, Wahl\'en, 2016)\cite{ehrnstrom2016whitham}.  Vargas-Maga{\~ n}a et al. \cite{11}  reported that Whitham-Boussinesq type systems, which are far simpler than the full water wave equations, can be used to accurately model surface undular bores, this is a  generic type of wave phenomenon arising as solutions of nonlinear dispersive wave equations that consists of two edges, the trailing edge and leading edge propagating with different speeds with a modulated dispersive wave-train between these. J. Carter and collaborators arrived at similar conclusions in \cite{carter2018bidirectional,carter2021fully} when comparing numerical results with experiments in a water wave tank. 
Many questions remain open around the  existence of solitary waves, the computation of these waves in a finite and variable depth and and their interactions such as the ``head-on'' and ``overtaking'' type collision in flat, variable and beach topography. In this research we want to explore using  careful simulations the pairwise solitary wave interactions and the Lax categories within the context of the two equations of the Whitham-Boussinesq type models


In the first part of this work we construct numerically solitary wave solutions of two Whitham-Boussinesq type models
 using an iterative Newton-type  method. For this purpose, we first write the Whitham-Boussinesq systems in the moving reference frame determined by the phase velocity of a traveling wave-type solution. Then we write the system of equations in the Fourier space, with this approach Whitham-Boussinesq systems are approximated via a finite dimensional nonlinear equation system that is solved using an iterative Newton-type method. We pointed out that the equations involves a nonlocal operator  that is  a pseudo-differential operator. This method has been previously implemented and adapted in other dynamical systems of interest to physics and on all cases has yielded a precision level of up to $10^{-16}$ digits of accuracy. Successful examples and effective implementation of the method on unidirectional nonlocal wave equations can be found in \cite{8,9,10}.


With this theoretical framework we obtained branches of solitary waves from both Whitham-Boussinesq type models. We carried out numerical quantitative and qualitative analysis on them that consists of analyzing the shape of solitary waves for various amplitudes, for high amplitudes solitary waves appear to have a non smooth limiting profile.
 We computed the speed of propagation of an exhaustive set of solitary waves for each  system, the limit wave that is reached by both systems fit a peakon solution in each case. An important aspect derived from this numerical analysis is that we obtained with a high level of accuracy the amplitude/velocity relationship for both systems, such "dispersion relationship" is a very useful description that relates two important concepts related to phase velocity and group velocity. The first tells us how fast a point on the wave with constant phase is moving, while the group speed measures how fast the wave energy is moving. An algebraic expression can be derived analytically for water wave models such as the Korteweg de-Vries equation, the Whitham equation, the Boussinesq models and many more systems, but it is not the case for the Whitham-Boussinesq models that we study in this article. This research reveled relevant information on the relationship of amplitude speed numerically for the solitary waves of these systems. The data obtained agrees with the results obtained with the "Dispersive Shock Fitting Method" and reported in \cite{11}, 
 It is important to highlight that the quantitative and qualitative knowledge of these solutions for these systems, as well as the relationships that they maintain between their macroscopic quantities of solitary waves is highly relevant for verifying the application of these systems in regimes of interest for Oceanography and Coastal Engineering.
 

 The second part of this research consists of the systematic and exhaustive study of the interaction of solitary waves, we will focus on the "overtaking" type collision of solitary waves for these systems at constant depth. The overtaking interaction is often referred to as the strong solitary wave interaction due to the relative importance of nonlinearity \cite{12}. It is known that for the KdV equation, a weakly nonlinear standard long-wave model of unidirectional waves that there are three categories of overtaking collisions classified by Lax \cite{12}. Lax gave geometric details of the interaction between two solitons  according to the number of local maxima during the collision into three categories. Furthermore, he showed that the classification only depends on the ratio of the initial amplitudes of the two solitons, so an algebraic classification was also determined. Each category corresponds to a different geometry of interaction of the colliding solitary waves. Several authors have investigated the collision of solitons experimentally and numerically with the idea of finding a geometric and algebraic classification similar to the interaction of two solitons described by Lax. These categories have been observed in multiple experiments and numerical models, so it is difficult to mention all the contributions. We summarize here some important results. Regarding the experimental works. Maxworthy \cite{17} they considered the collision of solitons in a tank full of water. Mirie and Su \cite{18} they used numerical methods to verify the Lax classification for a higher order model. More recently these categories have been observed in fully non-linear water wave models (Euler and Serre equations), see \cite{13,14,15} and a viscous core annular flow model \cite{16}.
 
 
 As a result of these investigations and the diversity of systems studied to this date, we pointed out that the qualitative characterization of the pairwise solitary wave interaction in many water waves systems is richer than the three algebraic categories of two-soliton interaction stated by Lax for KdV equation. It has been reported in \cite{9,15,16} the following:

 \begin{itemize}
\item[(i)] The type of interaction depends on the waves amplitudes, rather than
only their ratio, 
 \item[(ii)] The unimodal/bimodal type of interaction is more robust than the mixed category,
 \item[(iii)]The transitions between categories when the algebraic Lax classification is possible differ in each weakly nonlinear water wave model,
\item[(iv)] For some systems, as is the case of the Serre equations, a different category is reported than those of Lax, which they call the transition category.
\end{itemize}

The project carries out a systematic study of the pairwise solitary wave interactions and the Lax categories within the context of the two equations of the Whitham-Boussinesq type models using careful numerical simulations, we find that the geometric characterization for the Whitham-Boussinesq models is closely related to the three Lax categories of two-soliton collisions for the KdV equation. Nevertheless, we reveal interesting differences addressed in the facts mentioned above for each of the Whitham-Boussinesq systems presented in this work, such as the robustness and the instability of some categories and the range of transitions.

 The organization of the article is as follows. In section 2, we formulate
the Whitham-Boussinesq systems, referred to as System HP and System ASMP. In section 3 we present the iterative Newton-type method   to find the solitary waves of these Whitham-Boussinesq type systems and  study their propagation.
  The results on a qualitative and numerical analysis of solitary waves  governed by each of these Whitham Boussinesq systems and  on the pairwise solitary wave interactions and the Lax categories within the context of these systems using careful numerical simulations are presented in section 4.

{
{\section{Whitham-Boussinesq models}}
 \color{black}

{We consider bidirectional long-wave models with an emphasis on fully dispersive models such as bi-directional non-local extensions of Boussinesq models. The Boussinesq approach assumes waves whose wavelength is much longer than the water depth. This implies that the parameters $\epsilon=a/ h_0$ and $\delta=h_0^2 / l^2$ are small. Here,  $l$  is the typical wavelength, $a$ is the typical surface elevation and $h_0$ is the mean depth, see {Fig. \ref{fig:2}}.
\begin{figure}[h!]
\begin{centering}
\includegraphics[scale=0.48]
{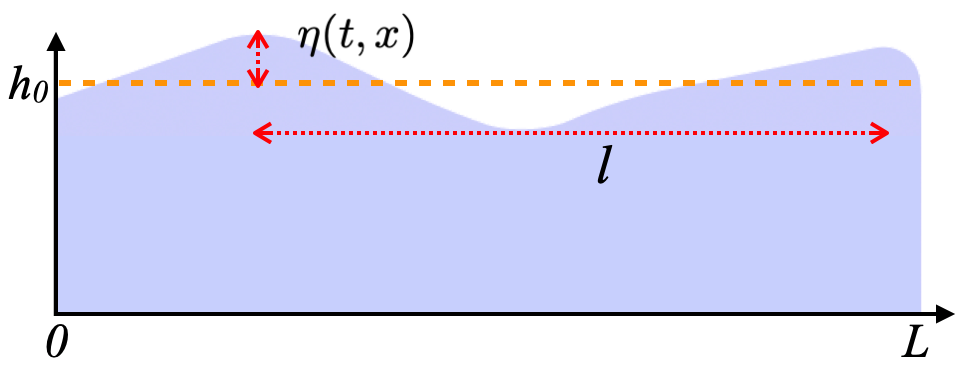}
\vspace{-5.0mm}
\caption{\small{Schematic illustration of the \textit{shallow water waves} (Boussinesq approach).
}}
       \label{fig:2}
       \end{centering}
\end{figure}

Boussinesq systems do not exhibit short wave effects, such as breaking and peaking, which are exhibited by solutions of the full water wave equations. To explore short wave effects in the context of weakly nonlinear dispersive wave equations, Whitham proposed replacing the third-order dispersion of the Korteweg-de Vries equation (unidirectional Boussinesq equation) with full water wave dispersion, resulting in what is now termed the Whitham equation \cite{whitham1967variational}.

Following this heuristic leads to the derivation of the Whitham-Boussinesq models that we introduce below.
\vspace{-0.0mm}
\begin{itemize}
    \item[]\textbf{System  I.}  (HP-model) This Whitham-Boussinesq system has been derived from a combination of the full dispersion relation for water waves and the convective term of the nonlinear shallow water equations.  The system has been termed the ``full-dispersion shallow water equations'' \cite{hur2018wave,hur2019modulational}.
\begin{eqnarray} \label{WBC}
\nonumber
\eta_t &=& -u_{x}-  (\eta u)_x,  \\ 
 u_t &=&- uu_{x}  -\partial_{x}\left(\left[ \frac{\tanh
 h_0D}{h_0D}\right] \eta\right).
\end{eqnarray} 
where the horizontal velocity $u=\xi_x$, with $\xi$ the hydrodynamic potential in the free surface, $D=-i\partial_x$ and $h_0$ is the mean depth of the fluid. 
 The system is Hamiltonian \cite{hur2019modulational} with Hamiltonian
 \stepcounter{equation}
 \begin{equation}\label{hamflat}
H = \frac{1}{2} \int_{\mathbb{R}} \left(\xi_x^2 + g \eta \left[\frac{\tanh h_0D}{h_0D}\right]\eta + \eta(\xi_x)^2 \right)dx.
\end{equation}
\vspace{-6.0mm}

\item[]{}\textbf{System II.} (ASMP-model)This Whitham-Boussinesq system has been derived from the Hamiltonian formulation of the water wave equations \cite{aceves2013numerical}:
\begin{eqnarray} \label{WBD}
\nonumber
\eta_t &=& -\partial_{x}\left(\left[\frac{\tanh h_0D}{h_0D}\right] u\right)- (\eta u)_x,  \\ 
 u_t &=& -\eta_x - uu_{x}.
\end{eqnarray} 
\end{itemize} 
Its Hamiltonian is 
\vspace{-2.0mm}
 \begin{equation}\label{hamflat}
H = \frac{1}{2} \int_{\mathbb{R}} \left( \xi D \tanh(h_0D) \xi + \eta(\xi_x)^2+ g\eta^2 \right)dx.
\end{equation}

In what follows we denote the nonlocal operator  $ \mathcal{K}:=\left[\frac{\tanh h_0D}{h_0D}\right]$ and the mean depth is normalized as  $h_0=1$.

One crucial difference between both Whitham-Boussinesq systems is that the  Hamiltonian of System II is obtained using a variational approach from the Hamiltonian of the water wave problem \cite{zakharov1968, stability, craig1993numerical} and this Hamiltonian corresponds to the total energy (the sum of {potential} and {kinetic} energy) of the water  mass.

We refer to this system (\ref{WBC}) as \textbf{HP- Model} and to  system (\ref{WBD}) as \textbf{ASMP-Model}. Both systems are bi-directional, have the same linear phase velocity as the Euler equations and have the same dispersion-less system associated to them.

In the following section we present the numerical methods to compute solitary waves for systems (\ref{WBC})-(\ref{WBD}) and their evolution over time.

\section{Numerical methods}
The  Whitham-Boussinesq systems (\ref{WBC}) and (\ref{WBD}) are solved numerically using a Fourier pseudospectral method. Since the method to solve both systems are similar we present the details only for system I. We make use of the Fourier transform to write system  (\ref{WBC}) in the Fourier frequency space as
\begin{align}\label{MC1}
\begin{split}
\hat{\eta}_{t} - ik\hat{u}+ ik\widehat{\eta u}=0, \\
\hat{u}_{t} - \frac{ik}{2} \widehat{u^2}-\frac{\tanh(k)}{k}\hat{\eta}=0.
\end{split}
\end{align}
Equations (\ref{MC1}) are solved  with a uniform grid with  even points $N$ in a periodic computational domain $[-L,L]$. The spatial grid is defined
\begin{equation}\label{grid}
x_j = -L + (j-1)\Delta x, \mbox{ $j=1,2,\dots$, N, where $\Delta x = 2L/N$.}
\end{equation}
In order to have high accuracy, we compute the Fourier transforms and spatial derivatives spectrally  using the Fast Fourier Transform (FFT) \cite{Trefethen:2000}.  The initial data is evolved in time through the classical Runge-Kutta fourth-order method (RK4) with time step $\Delta t$. 

Whitham-Boussinesq solitary waves  with speed $c$,  amplitude velocity $A$ and with its crest located at $x=0$ are computed through an iterative Newton's type method by solving the equations
\begin{align}\label{Newton}
\begin{split}
{\eta}= \mathcal{K}^{-1}*\Big[c{u}_{x} + \frac{1}{2} (u^2)_{x}\Big], \\
-c{\eta}_{x} - u+ (\eta u)_{x}=0.
\end{split}
\end{align}
On the grid points defined in equation (\ref{grid}), we denote by $u_j=u(x_j)$, $u_{x,j}=u_x(x_j)$,  $\eta_j=\eta(x_j)$, $\eta_{x,j}=\eta_x(x_j)$ and $(\eta u)_{x,j}=(\eta u)_x(x_j)$. The discretised version of system ({\ref{Newton}})  gives rise to a system of ($N+1$) equations  with ($N+1$) unknowns
\begin{align}\label{EE3}
\begin{split}
G_{j}(x_1,x_2,...,x_{N},c):=-c{\eta}_{x,j} - u+ (\eta u)_{x,j}=0,  \mbox{ for $j=1,2,\dots N$,} \\
G_{N+1}(x_1,x_2,...,x_{N},c):=u_{N/2+1}-A=0.
\end{split}
\end{align}
The discretization chosen allow us to compute all spatial derivatives and the nonlocal  operator $\mathcal{K}$ in equations (\ref{Newton}) with spectral accuracy in Fourier space through the FFT \cite{Trefethen:2000}. The  Jacobia's system  for the Newton iteration is found by finite differences in the unknowns and the stopping criterion is
\begin{align*}
\begin{split}
\frac{\sum_{j=1}^{N+1}|G_{n}(x_1,x_2,...,x_{N},c)|}{N+1}< \delta,
\end{split}
\end{align*}
where $\delta$ is a  given tolerance. In all simulations, the initial guess considered $(u_0,c_0)$ has KdV-soliton profile as follow 
\begin{equation}\label{Guess}
u_{0} = A\sech^{2}\Bigg(\sqrt{\frac{3}{4}A}x\Bigg) \mbox{ and } c_0 = -\frac{A}{2}.
\end{equation}
where $a_0=0.1$. The solution is then continued in the amplitude and speed by using the prior converged solution of the Newton's method as the initial guess to obtain a new solution with a larger amplitude and new speed.}

\section{Numerical results}

{ \subsection{Solitary waves on Whitham-Boussinesq type models}

There is not an exact an explicit expression of this kind of traveling wave for these systems, this is due to the nonlocal operator describing the evolution equations of the wave profile in the case of \textbf{Model ASMP}  and of the hydrodynamic surface potencial in the case of \textbf{Model HP}.  Recently there has been a few published attempts in describing numerically solitary-waves on Whitham-Boussinesq type systems and revealing their general features, \cite{Dinvay2021, Linaresetal}. But there are still  many open questions around these traveling waves for these weakly nonlinear, nonlocal systems and a systematic and more exhaustive approach is required in order to get general facts on the shape profiles of these waves and on qualitative and quantitative features of these traveling waves associated to \textbf{Model HP} and \textbf{Model ASMP}.  


In this section we present branches of solitary waves governed by Whitham-Boussinesq-type models that has been constructed with an iterative Newton-type and continuation method  described in previous section. We carried out numerical quantitative and qualitative analysis on these branches of solitary waves revealing interesting and specific physical characteristics.  In figure \ref{fig:2}a and \ref{fig:2}b we are showing the solitary waves profiles for a large range of amplitudes for \textbf{Model HP} and for \textbf{Model ASMP} respectively. We can appreciate that for higher amplitudes, solitary waves appear to have a non smooth limiting profile. The limit wave reached by both systems is singular and fits a peakon solution at each case. Figures \ref{fig:4}b and \ref{fig:4}b, shows a  perfect  fitting of the solitary wave closes to the highest amplitude reached in the branch with the associated  Whitham-peakon profile described by the solutions $f(x)=1.5953e^{-\frac{\pi}{1.5}|x|}$ and $f(x)= 0.4303e^{-\frac{\pi}{3.5}|x|}$ presented in figure \ref{fig:4}a and figure \ref{fig:4}b for \textbf{Model HP} and for \textbf{Model ASMP} respectively. Comparions between solitary waves of models textbf{Model HP}  and \textbf{Model ASMP} with KdV solitary waves are displayed in figures \ref{fig:3}a and \ref{fig:3}b respectively.


 We computed the speed of propagation of  a selection of representative elements belonging to the branch of solitary waves found for each  system. An important aspect derived from this numerical analysis is that we obtained with a high level of accuracy the amplitude/velocity relationship for both systems, such "dispersion relationship" is a very useful description that relates two important concepts related to phase velocity and group velocity. The first tells us how fast a point on the wave with constant phase is moving, while the group speed measures how fast the wave energy is moving. An algebraic expression can be derived analytically for water wave models such as the Korteweg de-Vries equation, the Whitham equation, the Boussinesq models and many more systems, but it is not the case for the Whitham-Boussinesq models that we study in this investigation \cite{11}. This research reveled relevant information on the relationship of amplitude speed numerically for the solitary waves of these systems. The data obtained agrees with the results obtained with the "Dispersive Shock Fitting Method" and reported in \cite{11}. 

First, we computed the Solitary waves of system I  to which the KdV solitary wave as initial guess was perfect to run the continuation Method. Then once we could get the solitary wave for system I we were able to use this wave as the initial guess for system II.
We did this in a systematic and exhaustive way in order to obtain a continum of solitary waves.
\begin{figure}[h!]
\begin{centering}
\includegraphics[scale=0.278]{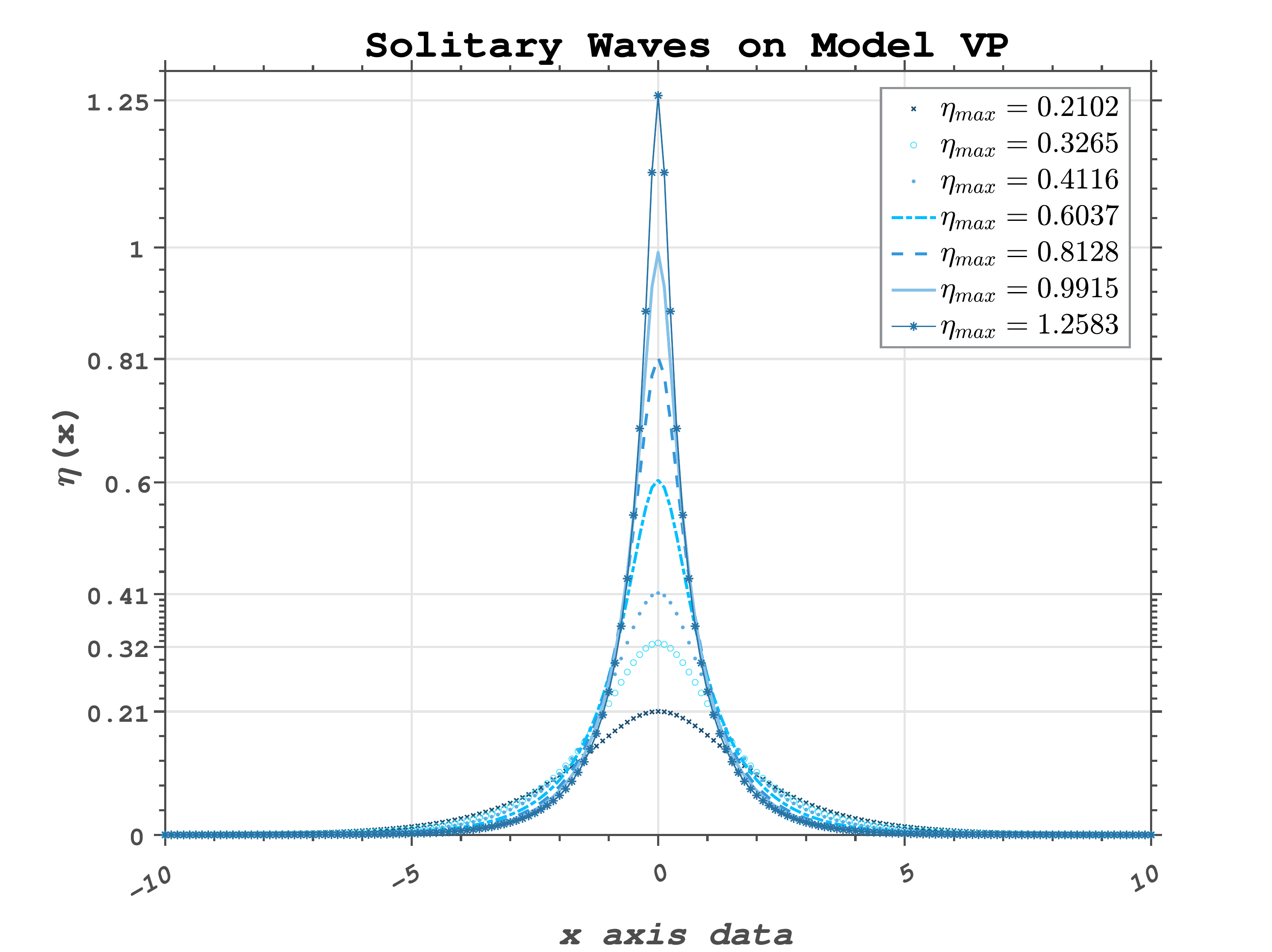}
\includegraphics[scale=0.291]{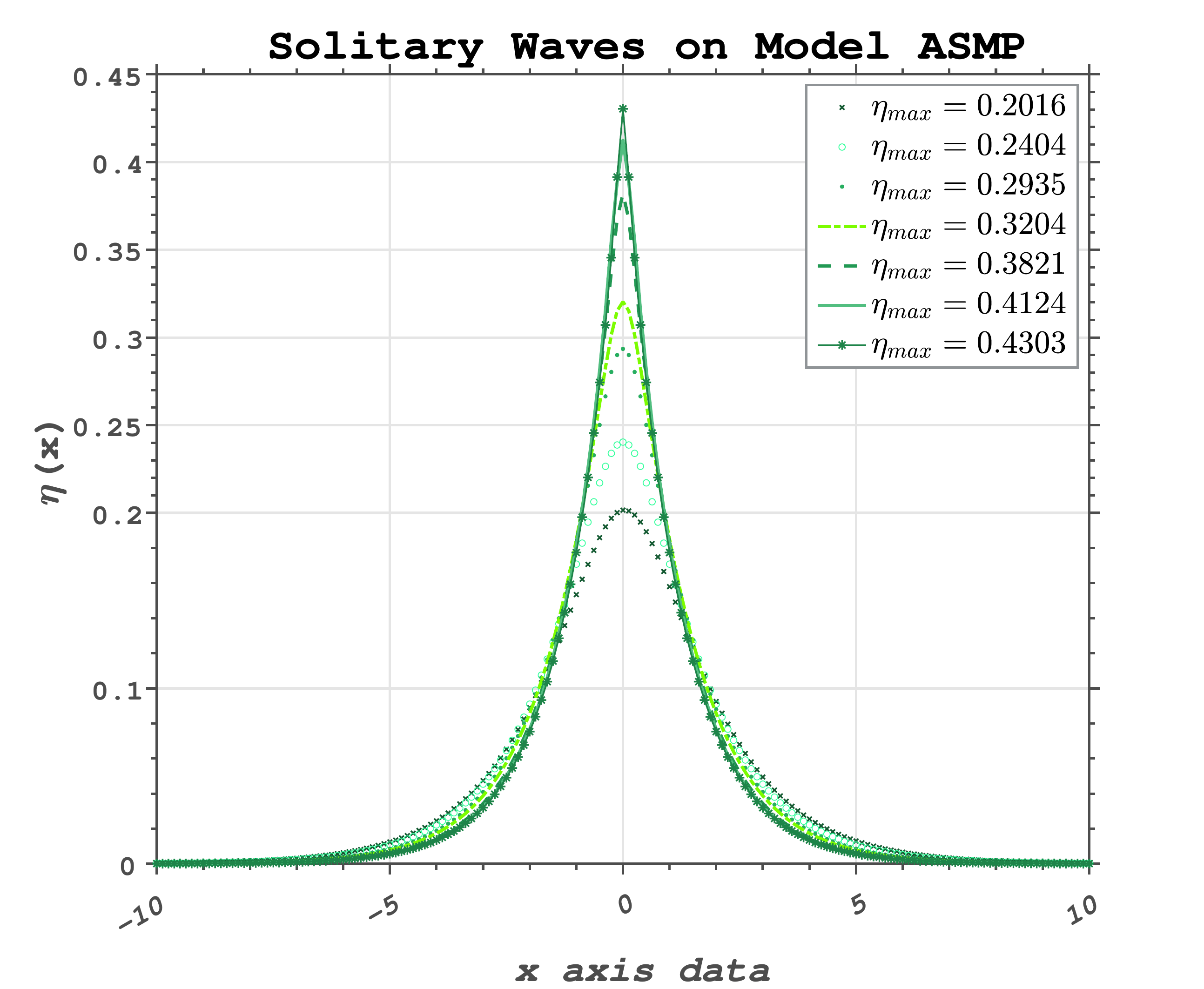}
\caption{\small{a) Solitary waves on system I. b) Solitary waves on system II }}
       \label{fig:2}
       \end{centering}
\end{figure}

Figures (2a) and (2b) shows some of these solitary waves for Model I and Model II respectively. They are displayed in a way that we can compare the wave profile of solitary waves of the same amplitude but that are moving with different speeds and also are evolving in a very different way, reaching similar limit states but at  different amplitude wave lengths, wave speeds and with distinct asymptotic regimes.  Solitary waves of system II appear to have a non smooth limiting profile. In the case of system I we captured solitary waves profiles of amplitude above 0.7 with our numerical continuation method.

 
To determine the amplitude of the solitary wave given the speed of the solitary wave, the amplitudeÐvelocity relation
for the solitary wave solution needs to be determined. This can be
found for the KdV, Boussinesq systems, but not for thecategorization
Whitham-Boussinesq equations. As far as the authors know it has not been a previous attempt in analyzing the amplitude-velocity relations for the solitary wave solutions of Systems I and II.

In the second part of this section que are obtained given all the data collected  from our previous analysis we were able to get an accurate amplitude-velocity relation of the solitary waves for these two systems.

The novelty in this plot is that shows the suitable agreement on the curves with the data collected from the DSW-fitting method we got from previos analysis on these systems that are out of the scope of this paper and the interested lector can consult the paper on undular bores on these systems \cite{11}.

We report an excellent  agreement in system II and Water wave model, referring to the wave amplitude between 0 to 0.27.
We can appreciate in the same figure that amplitude-velocity relation for system I is bending around 0.3 wave amplitude, that behavior has been previously observed in the amplitude/velocity relation collected from the DSW-fitting method applied to system I and system II for the  description of the macroscopic quantities describing the solitary edge limite of undular bores on system I and system II such as leading solitary speed and the amplitude measured numerically with the long time integration of a Riemann initial condition of system I and II that is resolved as a DSW.
\begin{figure}[h!]
\begin{centering}
\includegraphics[scale=0.256]{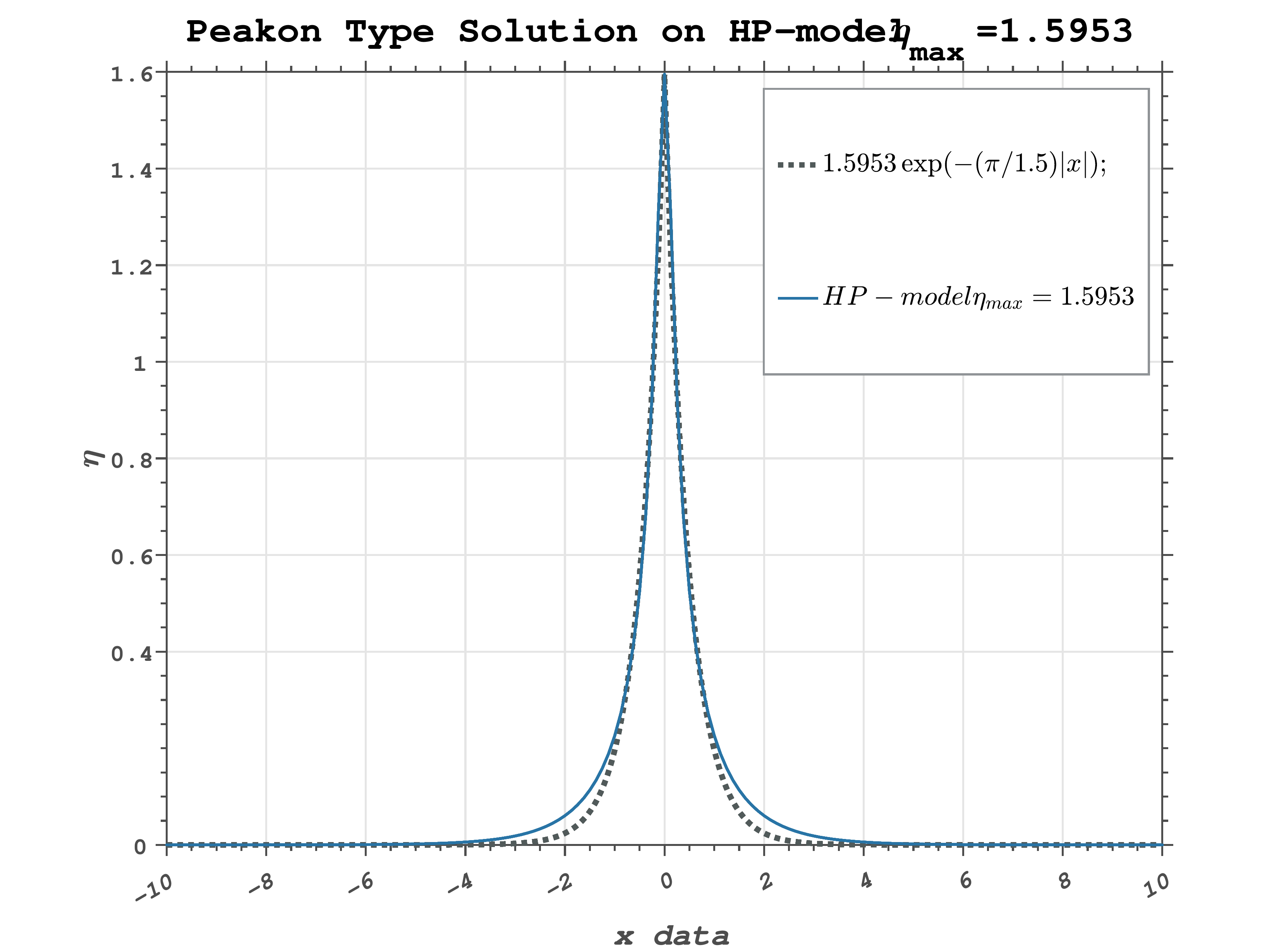}
\includegraphics[scale=0.26]{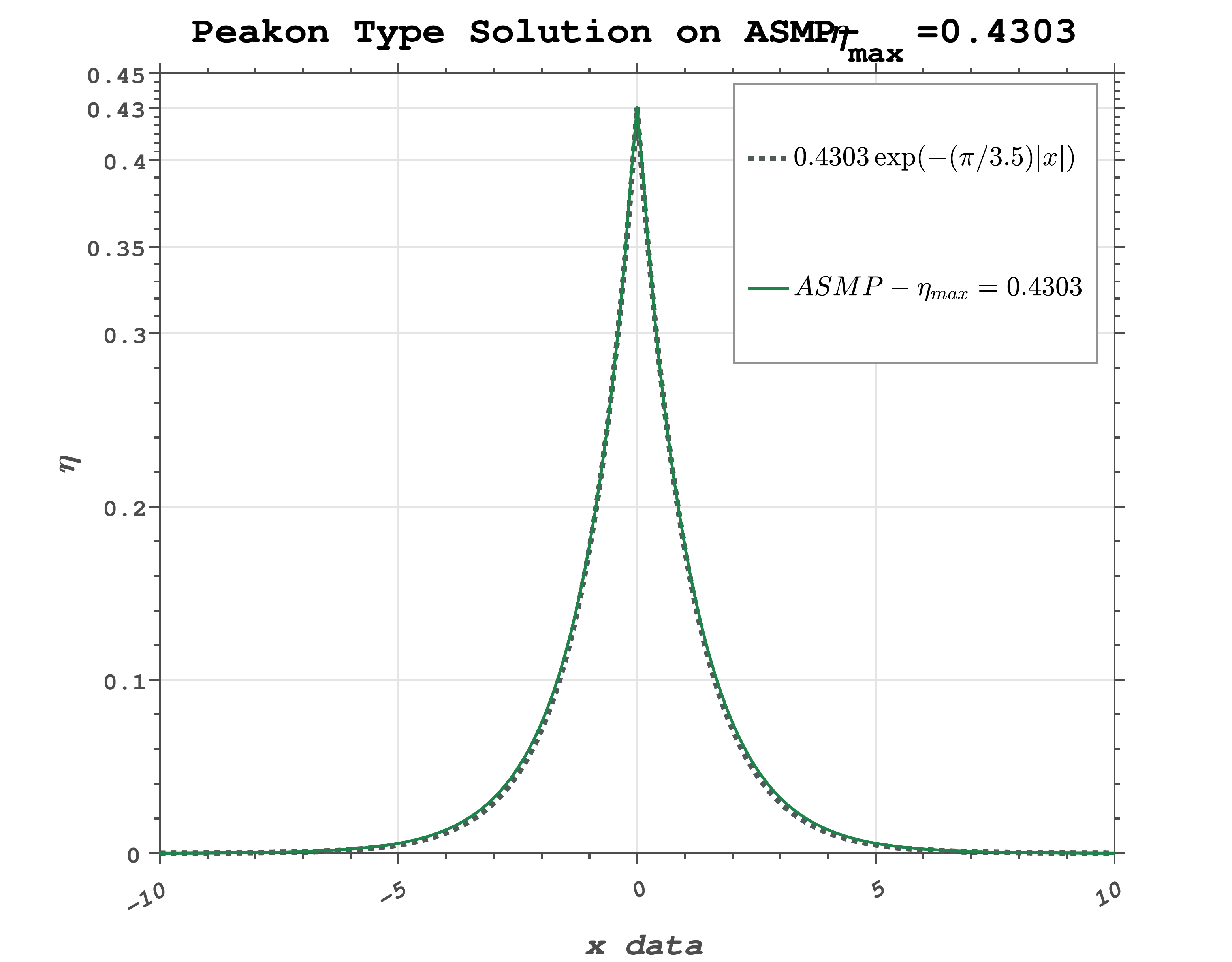}
\caption{\small{a) Perfect fitting of high amplitud solitary wave of \textbf{HP-Model}  with a Peakon-type solution. b)   Perfect fitting of high amplitud solitary wave of \textbf{ASMP-Model}  with a Peakon-type solution.}}
       \label{fig:4}
       \end{centering}
\end{figure}

\begin{figure}[h!]
\begin{centering}
\includegraphics[scale=0.27]{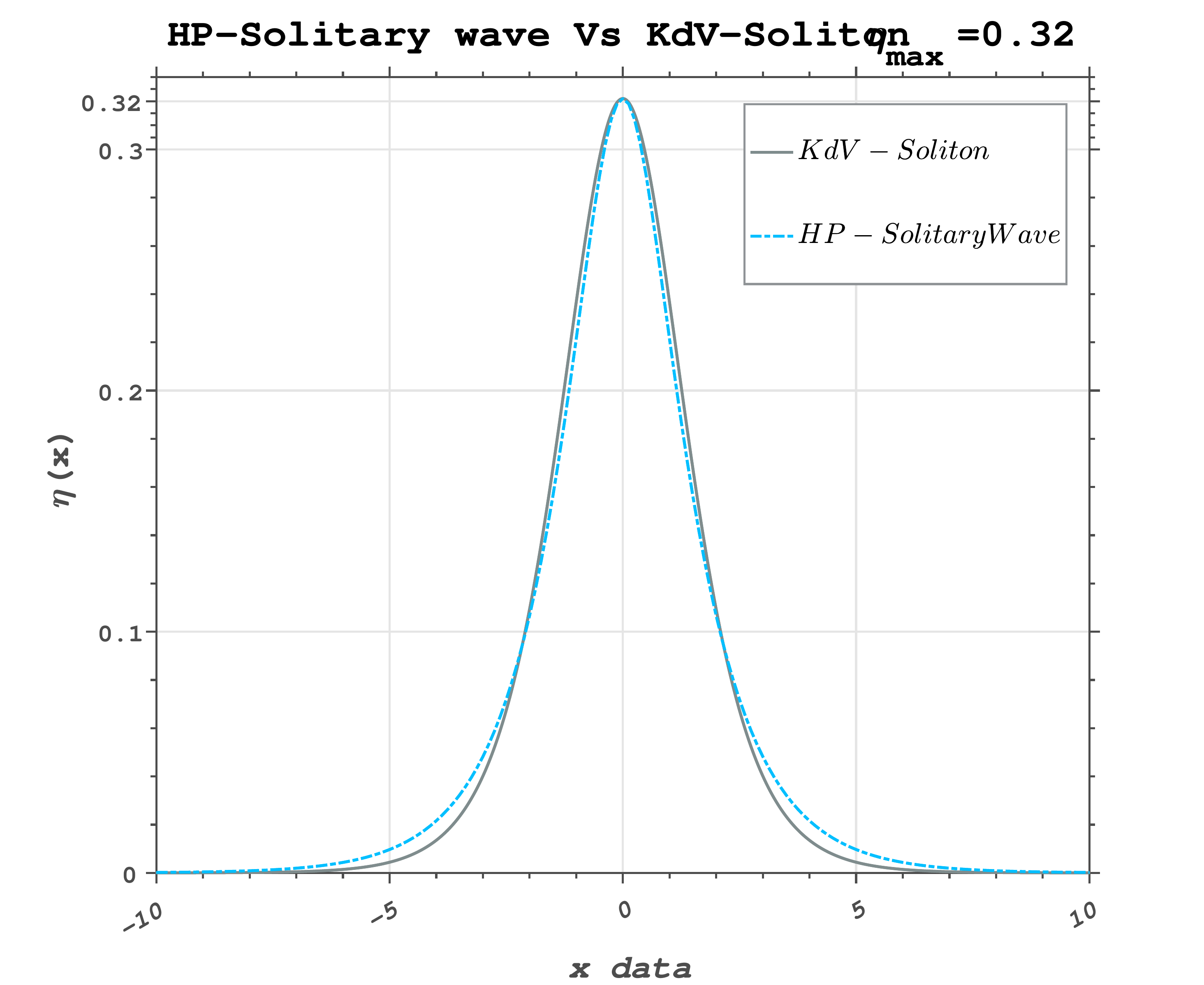}
\includegraphics[scale=0.28]{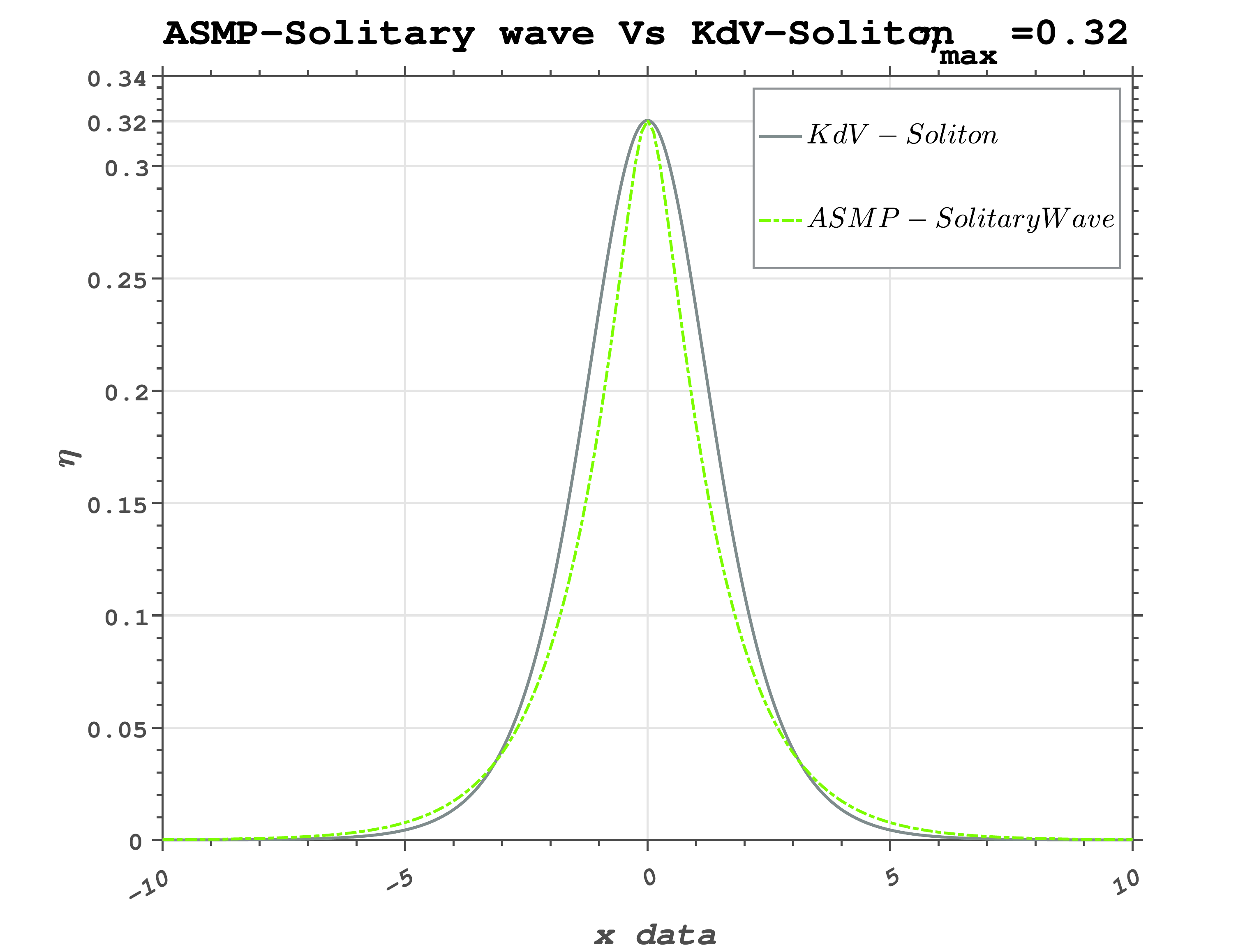}
\caption{\small{a) KdV-soliton given by \ref{Guess} with amplitude A=0.32 and solitary wave of \textbf{HP-Model} with the same amplitude. b) KdV-soliton given by \ref{Guess} with amplitude A=0.32 and solitary wave of \textbf{ASMP-Model} with the same amplitude}}
       \label{fig:3}
       \end{centering}
\end{figure}

\begin{figure}[h!]
\begin{centering}
\includegraphics[scale=0.48]{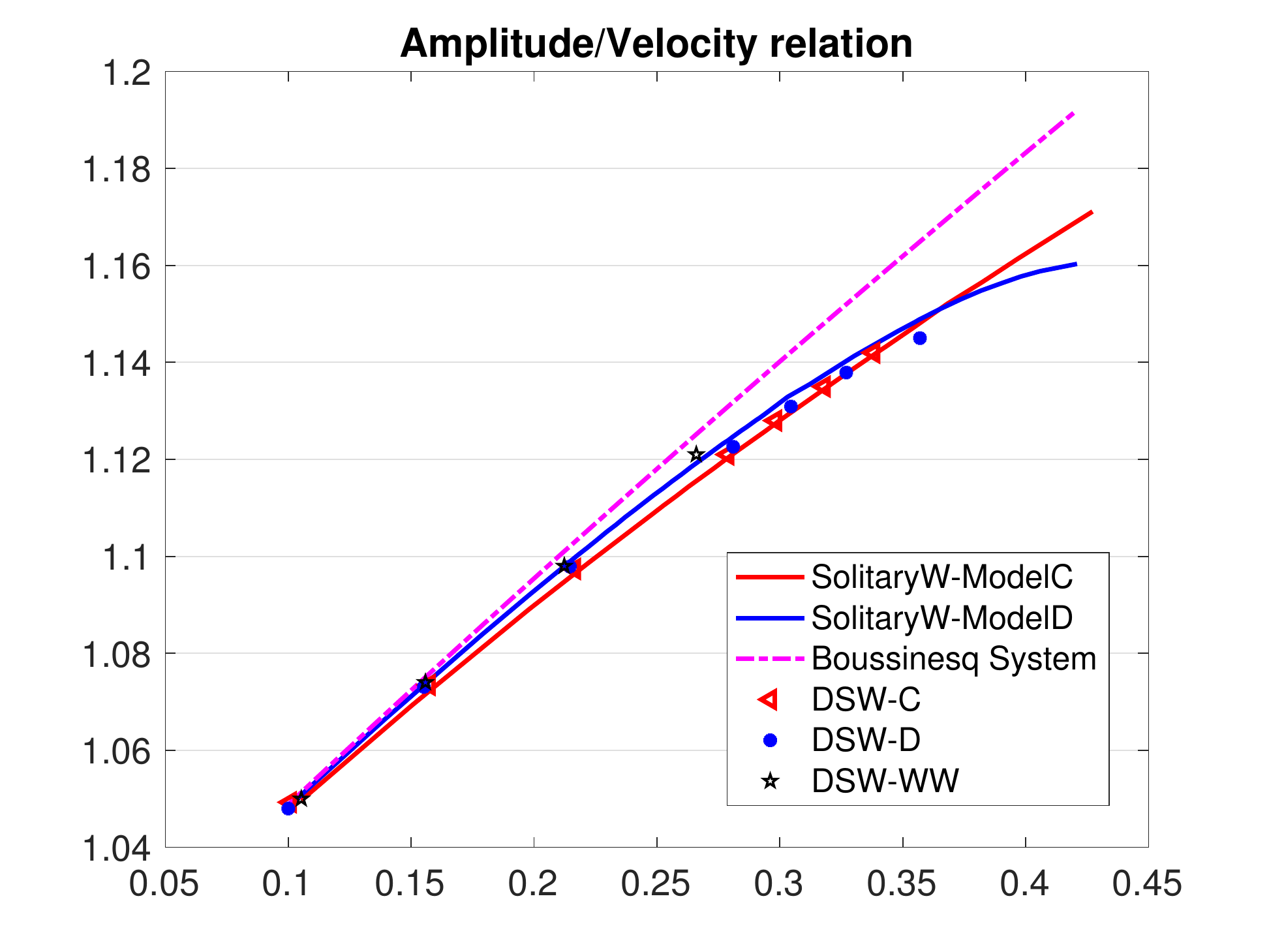}
\caption{\small{Amplitude-velocity relation for Boussinesq model, system I, system II and Water wave problem. The curves are created with data from the DSW-fitting method reported in \cite{11} and with the numerical solitary waves obtained with the Newton-type method.}}
       \label{fig:5}
       \end{centering}
\end{figure}

}

\subsection{Overtaking collisions of solitary waves}
We investigate collisions of two solitary waves for the Whitham-Boussinesq systems I and II in the same fashion as Flamarion and Ribeiro-Jr \cite{8}. To this end, we set two solitary waves computed through the iterative Newton-type method  (\ref{Newton}) far apart so  initially they have two well separated crests. We refer to  these two solitary waves by $S_1$ and $S_2$ and let  $A_1$ and $A_2$ be their respective amplitudes   with $A_1>A_2$. The initial data considered in all numerical simulations is
\begin{equation}\label{initial}
u(x,0) = S_{1}(x+10)+S_{2}(x-10).
\end{equation}

Now, we briefly recall the Lax classification for the two-soliton collisions for the KdV equation based on the ratio of the initial amplitudes as follows:
\begin{itemize}
	\item[{\bf (A)}] At any given time, there are always two-well defined separated crests, in other words, the solution has always two local maxima  if $A_{1}/A_{2}<(3+\sqrt{5})/2\approx 2.62$.	
	\item[{\bf(C)}] When $A_{1}/A_{2}>3$,  the number of crests changes according to the law as $2\rightarrow 1\rightarrow 2$ during the interaction. 
	
	\item[{\bf (B)}]  	This case has features of cases {\bf(A)} and {\bf(C)}. The number of local maxima varies acoording to the  law $2\rightarrow 1\rightarrow 2\rightarrow 1\rightarrow 2$.   This case happens  when $(3+\sqrt{5})/2<A_{1}/A_{2}<3$. 
\end{itemize}
It is worth to mention that the crest of the two solitons  are slightly shifted from the trajectories of the incoming centers after the collision takes place.

Our goal is to investigate whether or not the Whitham-Boussinesq systems satisfy the geometric and algebraic properties of the Lax-categorization. Secondly to report the features observed  after the collision takes place. It is worth to mention that the Whitham-Boussinesq systems are not integrable, thus right after the collision a small dispersive tail appears. Nonetheless  the two solitary waves have almost the same shape after their collision. 
 
 \subsubsection{Collisions for the Whitham-Boussinesq system I} 
We run a large number of collisions for solitary waves of system I. Based on our simulations we are inclined to say that the main difference for solitary wave collisions of Whitham-Boussinesq system I  is that the transitions occur for different values of the two incident amplitudes. We are able to find all three distinct geometric cases reported by Lax.  All three cases are displayed in great details in Figures  \ref{colisaoA}, \ref{colisaoC} and \ref{colisaoB}. Top figures display the collision of the two solitary waves and bottom pictures monitored the exact position of each crest throughout time  and the maxium local of the solution. Notice that right after the interactions during the collision the waves separate and travel almost as traveling solitary waves. Additionally, it seems that an algebraic classification of the collisions based on the amplitude of the incident solitary waves is also possible, see Table \ref{table1}. In this case the transition of categories occur in a different range from the one given for the KdV equation by Lax. More precisely, we have the following Lax-categorization for system I
\begin{itemize}
	\item[{\bf (A)}] At any given time, there are always two-well defined separated crests, in other words, the solution has always two local maxima  if $A_{1}/A_{2}<3.9$.	
	\item[{\bf(C)}] When $A_{1}/A_{2}>4.9$,  the number of crests changes according to the law as $2\rightarrow 1\rightarrow 2$ during the interaction. 
	
	\item[{\bf (B)}]  	This case has features of cases {\bf(A)} and {\bf(C)}. The number of local maxima varies acoording to the  law $2\rightarrow 1\rightarrow 2\rightarrow 1\rightarrow 2$.   This case happens  when $3.9<A_{1}/A_{2}<4.9$. 
\end{itemize}

A theoretical study on the boundaries that the transitions between two categories occur is needed and out of the scope of this work.  Since solitons do not have an exact expression for the Whitham-Boussinesq system I such study is very challenging. 

\begin{figure}[h!]
	\centering	
	\includegraphics[scale =1]{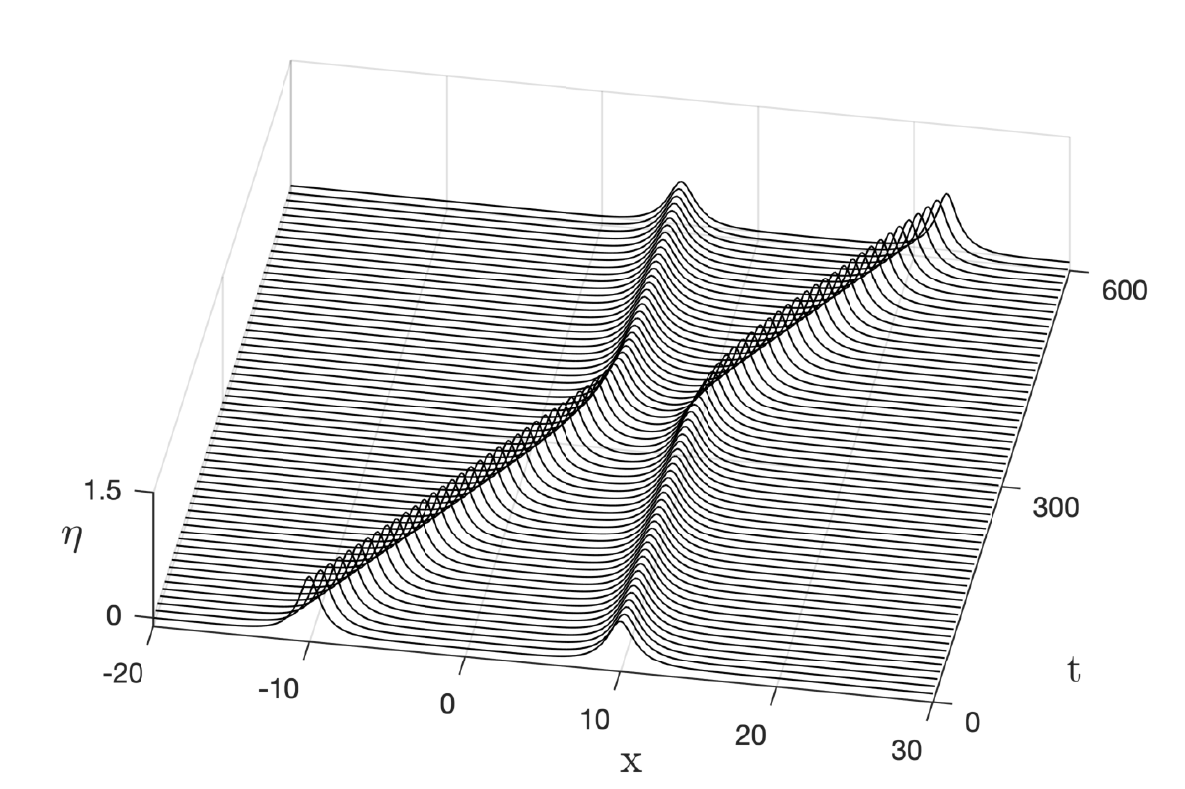}
	\includegraphics[scale =1]{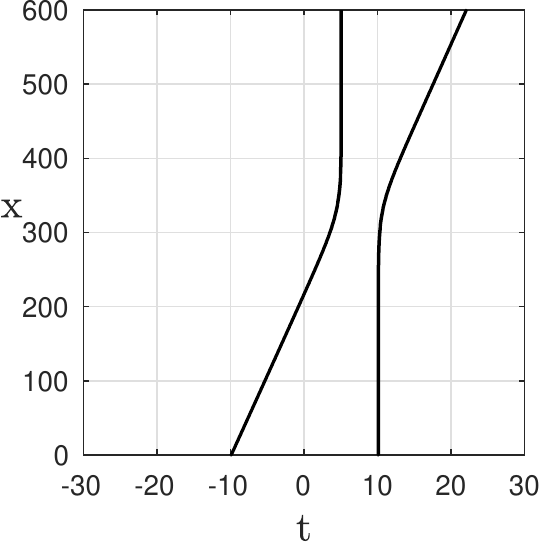}
	\includegraphics[scale =1]{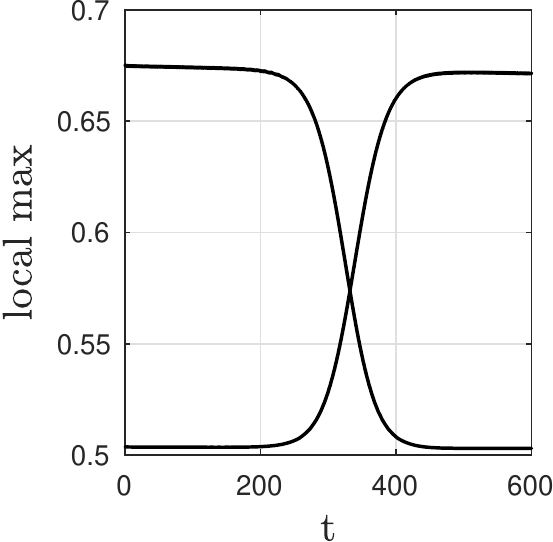}
	\caption{Top: Collision of two solitons for the Whitham-Boussinesq system C -- category {\bf(A)}. Bottom (left): Crest trajectories of the solitary waves before and after collision. Bottom (right): The local maxima of the solution as a function of time. Parameters $A_{1}=0.6754$, $A_{2}=0.5036$.}
	\label{colisaoA}
\end{figure}
\begin{figure}[h!]
	\centering	
	\includegraphics[scale =1]{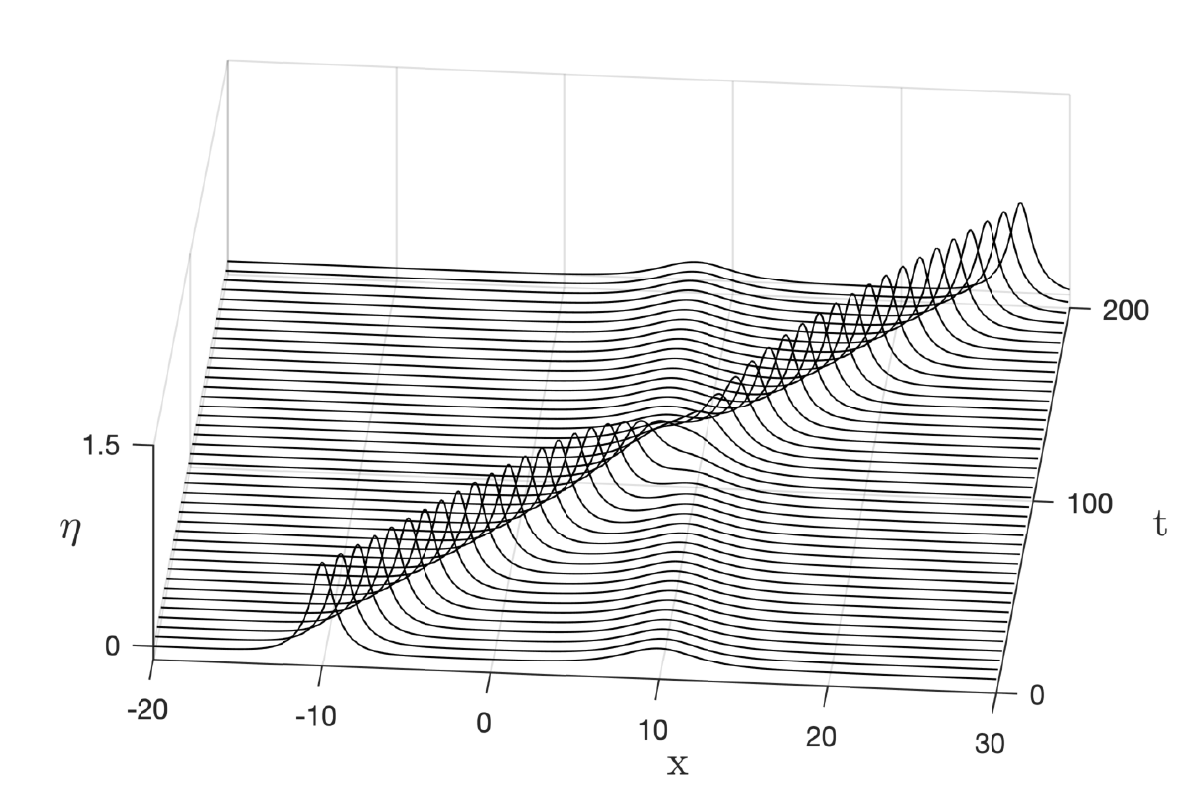}
	\includegraphics[scale =1]{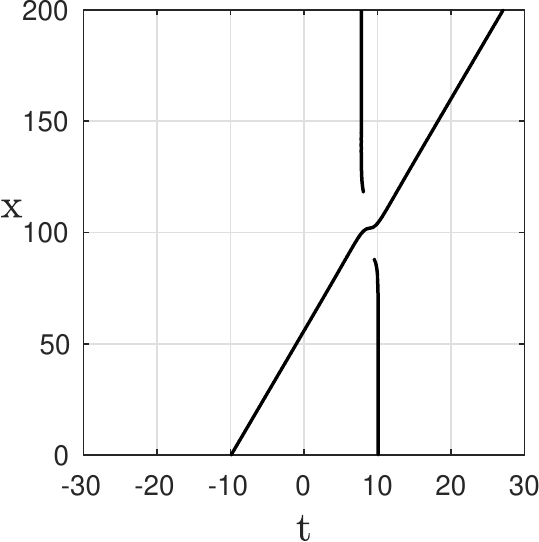}
	\includegraphics[scale =1]{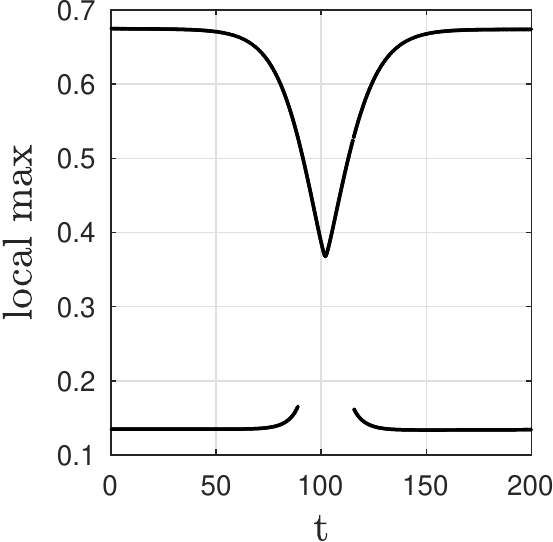}
	\caption{Top: Collision of two solitons for the Whitham-Boussinesq system C -- category {\bf(C)}. Bottom (left): Crest trajectories of the solitary waves before and after collision. Bottom (right): The local maxima of the solution as a function of time. Parameters $A_{1}=0.6754$, $A_{2}=0.1345$.}
	\label{colisaoC}
\end{figure}
\begin{figure}[h!]
	\centering	
	\includegraphics[scale =1]{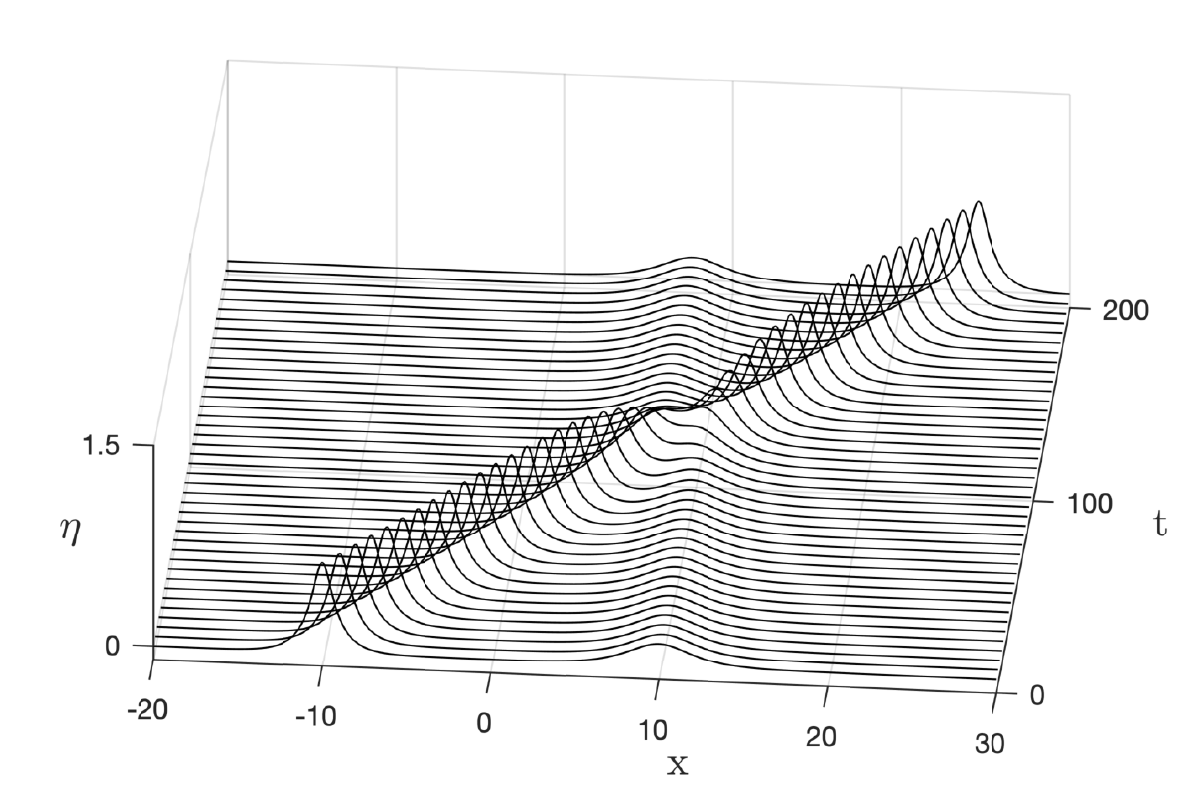}
	\includegraphics[scale =1]{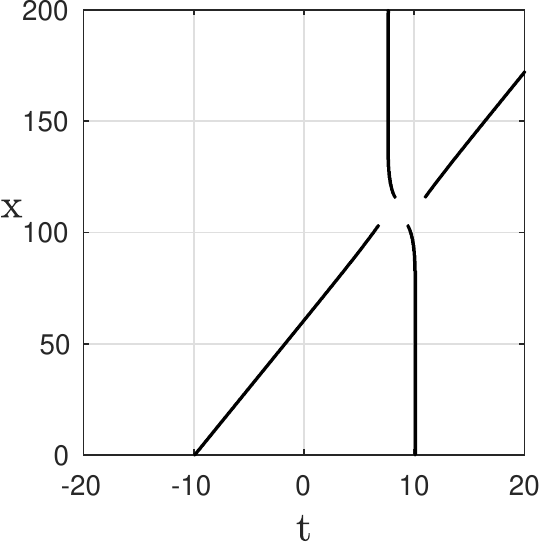}
	\includegraphics[scale =1]{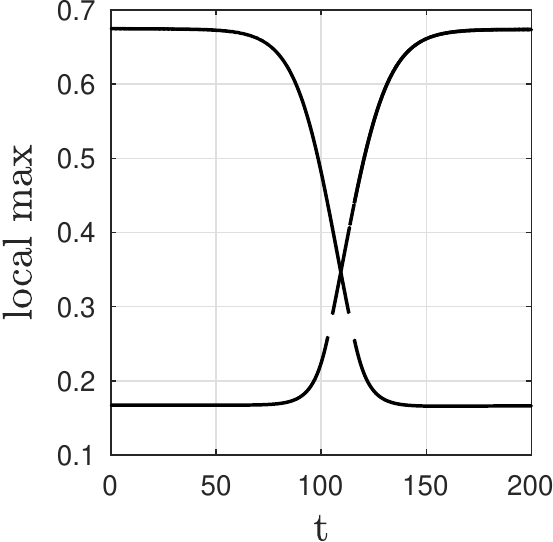}
	\caption{Top: Collision of two solitons for the Whitham-Boussinesq system C -- category {\bf(B)}. Bottom (left): Crest trajectories of the solitary waves before and after collision. Bottom (right): The local maxima of the solution as a function of time. Parameters $A_{1}=0.6754$, $A_{2}=0.1670$.}
	\label{colisaoB}
\end{figure}
\begin{table}[h!]
\centering
\begin{tabular}{c|c|c|c}\hline\hline
$A_{1}$ & $A_{2}$ & $A_1/A_2$ & Category   \\   \hline
0.6754 & 0.5864 &	1.1519 &	\bf{A} \\ \hline
0.6754 & 0.5036 &	1.3412 &	\bf{A} \\ \hline
0.6754 & 0.3542 &	1.9071 &	\bf{A} \\ \hline
0.6754 &0.3142  &	2.1497 &	\bf{A} \\ \hline
0.6754 & 0.2864 &	2.3587 &	\bf{A} \\ \hline
0.6754 & 0.2426 &	 2.7279&	\bf{A} \\ \hline
0.6754 & 0.1980 &	3.4111 &	\bf{A} \\ \hline
0.6754 & 0.1741 &	3.8806 &	\bf{A} \\ \hline
0.6754 & 0.1693 &	  3.9889&	\bf{A} \\ \hline
0.6754 & 0.1670 &	   4.0452 &	\bf{B} \\ \hline
0.6754 & 0.1658 &	   4.0738&	\bf{B} \\ \hline
0.6754 & 0.1634 &	   4.1324&	\bf{B} \\ \hline
0.6754 & 0.1506 &	 4.4840 &	\bf{B} \\ \hline
0.6754 & 0.1391 &	4.8554 &	\bf{B} \\ \hline
0.6754 & 0.1368 &	 4.9366 &	\bf{C} \\ \hline
0.6754 & 0.1345 &	5.0204 &	\bf{C} \\ \hline
0.6754 & 0.1323 &	5.1069 &	\bf{C} \\ \hline
0.6754 & 0.1277 &	  5.2885&	\bf{C} \\ \hline
0.6754 & 0.1164 &	  5.8004&	\bf{C} \\ \hline
0.6754 & 0.1052 &	 6.4146&	\bf{C} \\ \hline
\end{tabular}
\caption{Classification of the collision for different values of $A_1$ and $A_2$ for system I.}\label{table1}
\end{table}

 \subsubsection{Collisions for the Whitham-Boussinesq system II}

On the other hand, system II turn out to be very unpredictable. Although we are able to find all three type of collisions described by Lax in some cases, an algebraic classification of the collisions based on the ratio of the incident solitary waves is certainly not possible as it is displayed in Table \ref{table2}. As we can see the category changes form ${\bf A}\rightarrow {\bf C}\rightarrow {\bf B}\rightarrow {\bf C}$ as the ratio ($A_1/A_2$) increases.

For solitary waves of small amplitudes we expect similar results regarding the geometric Lax-categorization, however the range in which the category ({\bf B}) occurs seem to become thin. In particular when considering the largest solitary wave with amplitude $A_1=0.2467$ we are not able to find the category ({\bf B}). These results are different from the ones reported in the literature for the KdV equation \cite{12} and Euler equations \cite{13}.

\begin{table}[h!]
\centering
\begin{tabular}{c|c|c|c}\hline\hline
$A_{1}$ & $A_{2}$ & $A_2/A_1$ & Category    \\   \hline
0.3391 & 0.2467 &	   1.3746 &	\bf{A} \\ \hline
0.3391 & 0.2082 &	   1.6291 &	\bf{A} \\ \hline
0.3391 & 0.1621 &	   2.1035 &	\bf{A} \\ \hline
0.3391 & 0.1197 &	   2.8331 &	\bf{A} \\ \hline
0.3391 &  0.1008 &	    3.3634 &	\bf{A} \\ \hline
0.3391 &  0.1000 &	     3.3899 &	\bf{A} \\ \hline
0.3391  & 0.0998 &	     3.3969 &	\bf{C} \\ \hline
0.3391 &  0.0978 &	       3.4660 &	\bf{B} \\ \hline
0.3391  &  0.0799 &	    4.2434 &	\bf{C} \\ \hline
\end{tabular}
\caption{Classification of the collision for different values of $A_1$ and $A_2$ for system II.}\label{table2}
\end{table}

\begin{table}[h!]
\centering
\begin{tabular}{c|c|c|c}\hline\hline
$A_{1}$ & $A_{2}$ & $A_2/A_1$ & Category    \\   \hline
0.2467 &  0.2082 &	   1.1851 &	\bf{A} \\ \hline
0.2467 & 0.1612 &	  1.5303  &	\bf{A} \\ \hline
0.2467 & 0.1004 &	  2.4661  &	\bf{A} \\ \hline
0.2467 & 0.0978 &	     2.5215 &	\bf{A} \\ \hline
0.2467 & 0.0829 &	     2.9757&	\bf{A} \\ \hline
0.2467 & 0.0809 &	    3.0490 &	\bf{A} \\ \hline
0.2467 & 0.0799 &	   3.0870 &	\bf{C} \\ \hline
0.2467 & 0.0500 &	    4.9360 &	\bf{C} \\ \hline
0.2467 & 0.0300 &	    8.2245 &	\bf{C} \\ \hline

\end{tabular}
\caption{Classification of the collision for different values of $A_1$ and $A_2$ for system II.}\label{table3}
\end{table}



%
%
%
%
%
%

\end{document}